\documentclass{iaus}
\usepackage{graphicx}

\title[Mass loss and massive star evolution] 
{Mass loss and fate of the most massive stars}

\author[Jorick S. Vink]   
{Jorick S. Vink$^1$}

\affiliation{$^1$Armagh Observatory, College Hill, BT61 9DG, Armagh, 
United Kingdom \\ email: {\tt jsv@arm.ac.uk}}

\pubyear{2012}
\volume{279}  
\pagerange{xxx--yyy}
\jname{Death of Massive Stars: Supernovae and Gamma-Ray Bursts}
\editors{P. Roming, N. Kawai \& E. Pian, eds.}
\begin{document}

\maketitle

\begin{abstract}
The fate of massive stars up to 300$M_{\odot}$ is highly uncertain. 
Do these objects produce pair-instability explosions, or normal Type Ic 
supernovae? In order to address these questions, we need to know their 
mass-loss rates during their lives. 
Here we present mass-loss predictions for
very massive stars (VMS) in the range of 60-300$M_{\odot}$.
We use a novel method that simultaneously predicts the 
wind terminal velocities $v_{\infty}$ and mass-loss rate $\dot{M}$ 
as a function of the stellar parameters: (i) luminosity/mass $\Gamma$, (ii) metallicity $Z$, and (iii) 
effective temperature $T_{\rm eff}$. Using our results, we evaluate the likely
outcomes for the most massive stars.
\keywords{stars: mass loss, stars: evolution, stars: Wolf-Rayet, supernovae: general}
\end{abstract}

\section{Introduction}

Mass loss is the decisive parameter for predicting final stellar masses and 
the types of supernova (SN) explosion. 
Do the most massive stars 
disrupt as pair-instability SNe (PISNs), or do they produce normal SNe Ic?
When does this occur in conjunction with a long gamma-ray burst (GRB)? 
Is low metallicity $Z$ simply in {\it favour} 
due to lower mass-loss rates $\dot{M}$, or is it even 
a {\it stringent} requirement?
Furthermore, the formation of intermediate mass-black holes (IMBHs) 
and the stellar black-hole mass distribution are determined 
by $Z$-dependent $\dot{M}$ (Heger et al. 2003; Eldridge \& Vink 2006).

Another relevant issue concerns the stellar upper-mass limit. 
Until recently many researchers accepted a 150$M_{\odot}$ limit. 
Crowther et al. (2010) recently agued for 
much higher luminosities -- with masses twice as high -- 
for the WNh objects in dense clusters. 
A potential issue with the Crowther et al. luminosities 
is that these objects are clustered, involving a non-negligible chance 
of photon pollution from line-of-sight objects.

We have found a new WNh star VFTS\,682 in 30\,Dor 
(Evans et al. 2011; Bestenlehner et al. 2011). It is 
a near-identical twin of one of the 'Crowther' stars, R136a3. 
Surprisingly, VFTS\,682 is in apparent 
isolation from the R136 cluster (see Bestenlehner et al. for a discussion 
on isolated formation or a ``slow runaway'' status). This   
enables a check on the reliability of 
the luminosities derived for the core stars. 
Our finding of log$(L/L_{\odot}) = 6.5 \pm 0.2$ 
for VFTS\,682 provides support for 
high luminosities and masses, as 
the chance of line-of-sight pollution is small 
for this isolated star. 
Mass-loss rates for VMS up to 300$M_{\odot}$
are needed to establish their fate.
VMS are extremely close to the Eddington limit 
$\Gamma$ $=$ $g_{\rm rad}/g_{\rm grav}$ $=$ $\kappa L/(4 \pi c G M)$.

\section{Method: Monte Carlo mass-loss predictions}

Stellar winds from massive stars are driven by 
radiation pressure on spectral lines (Castor et al. 1975, CAK), 
predominantly on Fe. 
The approach we use to compute $\dot{M}$ for VMS is similar 
to the Monte Carlo method used to predict $\dot{M}$ for 
normal OB stars (Vink et al. 2000).
Until 2008 our methodology was semi-empirical, as 
we assumed a velocity law that reached 
a certain empirical $v_{\infty}$. 
M\"uller \& Vink (2008) suggested a new line-force 
parametrization that explicitly depends on radius 
(rather than the velocity gradient, as in CAK theory). 
We predicted $v_{\infty}$ within $\sim$25\% of the observations. 
In Muijres et al. (2012) we tested the M\"uller \& Vink 
approach by comparison to hydrodynamical models. 
As both methods gave similar results, we 
use the M\"uller \& Vink approach for VMS.

Nugis \& Lamers (2002) and Gr\"afener \& Hamann (2008) 
studied radiative driving due to Fe-peak opacities in deep 
photospheric layers of Wolf-Rayet (WR) winds.  
As $\Gamma$ crosses unity in deep layers, the sonic point is located
at high optical depth, leading to the initiation 
of optically-thick winds. In the Monte Carlo models, one traces 
the driving over the entire wind, and as
the bulk of the energy is transferred in the supersonic portion of the wind, 
one is less susceptible to the details of the photospheric region. 
Our strategy allows us to explore the transition from optically thin 
O-star winds to optically thick WR winds. 

\begin{figure}
\begin{center}
 \includegraphics[width=3.4in]{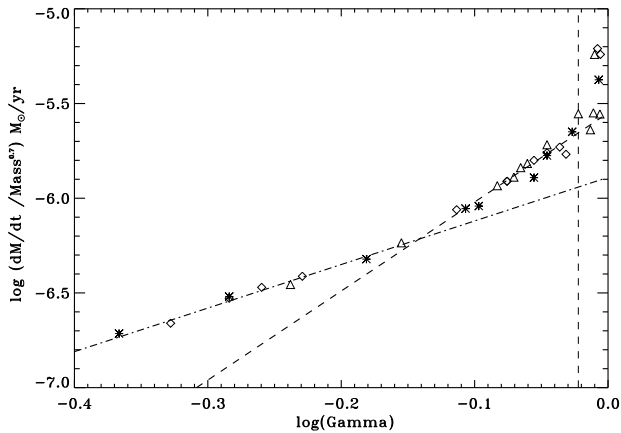} 
\caption{Mass-loss predictions versus the Eddington parameter $\Gamma$ -- divided by $M^{0.7}$. 
Symbols correspond to models of different mass ranges (Vink et al. 2011a).}
\label{f_mdotvink}
\end{center}
\end{figure}

\section{$\dot{M}-\Gamma$ dependence - Do PISNs exist at $Z_{\odot}$?}

In Figure\,\ref{f_mdotvink} we show mass-loss predictions for VMS 
as a function of the Eddington parameter $\Gamma$ 
(see Vink et al. 2011a for details). Most notable is the presence of 
a {\it kink} in the relation. 
For O-type stars with ``low'' $\Gamma$ and 
optically-thin winds, the $\dot{M}$ $\propto$ 
$\Gamma^{x}$ relationship is shallow, with $x$ $\simeq$2. 
There is a steepening at higher $\Gamma$, where
$x$ becomes $\simeq$5. Here the objects show optically thick 
WR-like winds, with optical depths and wind efficiencies above unity. 

Gr\"afener et al. (2011) recently provided empirical evidence for 
our predicted steep exponent ($x$ $\simeq$5), but 
note that there are still issues with our $v_{\infty}$ values 
for the high $\Gamma$ range. 
For now we employ the Vink et al. (2000) mass-loss recipe for 
our assessment of the fate of the most massive stars. These mass-loss rates agree 
extremely well with the rates discussed by Crowther et al. (2010) for the 
30\,Dor R136 core stars. 
We have recently also calibrated the Vink et al. rates using an 
analytic method and applied it to the most massive stars 
in the Arches cluster (Vink \& Gr\"afener 2012). 

Using Vink et al. (2000) rates for a star 
starting with 300$M_{\odot}$ we find $\dot{M}$ $= 10^{-4.2}$ 
$M_{\odot} {\rm yr}^{-1}$.
For a lifetime of 2.5 Myrs, this leads to a total 
main-sequence mass lost of $\simeq$150$M_{\odot}$. 
Additional mass loss during the core
helium WR phase should further``evaporate'' the object. Our results indicate
that there is little room for substantial additional mass loss in 
luminous blue variable (LBV) eruptions. 
Our results also imply that IMBHs and 
pair-instability explosions are unlikely. Unless we go to lower $Z$ environments.

\section{$\dot{M}-Z$ dependence - Are GRBs confined to low $Z$?}

The issue of mass loss and evolution at low $Z$
has gained attention due to the issue of the 
progenitors of long GRBs.
Within Woosley's collapsar model, GRB progenitors require two key properties: (i) 
a rapidly rotating core, and (ii) the absence of a hydrogen envelope. 
Therefore, GRB progenitors are thought to be rotating WR stars. 
The potential problem with this is that WR star have high mass loss which 
should remove the angular momentum before the core collapses. 

In the rapidly rotating stellar models of Yoon \& Langer (2005), 
the objects evolve ``quasi-homogeneously''. 
The stars are subject to a strong magnetic coupling between 
the core and envelope. If the rapid rotation can be maintained 
due to low main-sequence mass loss in low $Z$ galaxies, the objects may 
avoid slow-down in a red supergiant (RSG) or LBV phase, and directly 
become rapidly rotating WR stars. 
If the WR winds also depend on Fe driving (Vink \& de Koter 2005), the 
WR stars can maintain rapid rotation towards the very end, making GRBs -- 
but {\it only} at low $Z$. 

GRB data presented at this meeting suggest that GRBs are not restricted 
to low $Z$, but there seems to be a need for a GRB channel at high $Z$. 
We have recently identified a sub-group of rotating Galactic WR 
stars -- allowing for a potential solution to this problem 
(Vink et al. 2011b; Gr\"afener et al. 2012b). 
Spectropolarimetry surveys show that the majority of WR stars have spherically 
symmetric winds indicative of slow rotation, but a small minority display 
signatures of a spinning stellar surface. 
We found this spinning sub-group to be surrounded by ejecta 
nebulae, which are thought to be ejected during a recent RSG/LBV phase, which 
suggests that these WR stars are still young and rotating. 

If the core-surface coupling were strong enough, the cores would not be 
expected to rotate rapidly enough to make a GRB, but if the core-envelope 
coupling is less efficient, they may have the required angular momentum 
in their cores to make GRBs. 
In most high $Z$ cases these stars would nonetheless still be expected 
to spin down due to mass loss, but within our post-RSG/LBV 
scenario one would not exclude the possibility of a high $Z$ GRB. 
Yet, low $Z$ environments are still preferred due to weaker WR winds. 

\section{$\dot{M}-T_{\rm eff}$ dependence - Do Luminous Blue Variables (LBVs) explode?}

\begin{figure}
\begin{center}
 \includegraphics[width=3.4in]{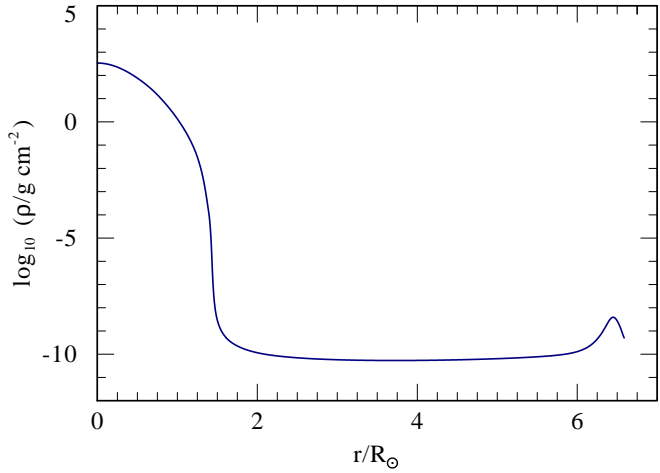} 
 \caption{Density vs. radius for a 23$M_{\odot}$ helium model of Gr\"afener et al. (2012a) showing
a density inversion. This leads to an inflation of the outer envelope.}
   \label{f_densinv}
\end{center}
\end{figure}

The stellar winds of O supergiants are fast ($\simeq$2000-4000 km/s) and 
transparent, whilst those emanating from lower $T_{\rm eff}$ 
B supergiants are much slower ($\simeq$100-1000 km/s). 
This is because O star winds are driven by high Fe ionization states, whilst those of B and later sub-types 
are driven by lower ones. This is wind bi-stability (BS).

LBVs increase their radii continuously on timescales of $\sim$10 yrs. These 
S\,Dor excursion across the HR diagram 
lead to winds with variable $v_{\infty}$ and $\dot{M}$. 
If the LBV wind changes instantaneously at the BS-jump, we can explain 
the double-throughed H$\alpha$ absorptions seen in LBV spectra (Groh \& Vink 2011). 
Intriguingly such double-throughed H$\alpha$ line profiles have also 
been seen in the luminous IIn SN 2005gj, which 
was for this reason suggested to have an LBV progenitor (Trundle et al. 2008). 
The same BS jump was also used to first suggest the LBV-SNe II link 
(Kotak \& Vink 2006). 

Even if $\dot{M}$ varies as a result of LBV radius changes, we still do 
not understand {\it why} LBVs change their radii 
(see Vink 2009 for a recent review). One possibility would 
be that the sub-photospheric outer envelopes of the stars become ``inflated'' 
as a result of the proximity to the Eddington limit (see Fig.\,\ref{f_densinv}).
Ishii et al. (1999) first studied the outer envelope inflation from stellar evolution 
models, and in Gr\"afener et al. (2012a) we developped an analytic explanation for how such an 
envelope inflation would occur. We described the radial inflation as a function of a dimensionless parameter W, which largely 
depends on the topology of the Fe-opacity peak. For $W>1$, we discovered an instability limit 
for which the stellar envelope becomes unbound. 
Within our framework, we are in principle able to explain LBV S\,Dor 
variations. Stellar temperatures could be strongly affected, and there could be important
implications of the {\it radii} of progenitor WRs and LBVs prior to collapse, as 
SN with different sub-types II, SN Ibc, and GRBs.

\begin{discussion}

\discuss{Omukai}{300$M_{\odot}$ stars are unstable to pulsation by the epsilon mechanism. 
Did you include this effect in evaluating the mass-loss rate?}

\discuss{Vink}{No, we didn't. The epsilon mechanism is thought to grow too slowly, 
and is usually not considered all that relevant.}

\discuss{Omukai}{You said that the WR envelope has a density inversion during the inflation phase.
Is it hydrodynamically stable?}

\discuss{Vink}{The Gr\"afener et al. (2012) models are static, and until we 
have studied the hydrodynamic case we cannot be 100\% sure. However, 
the suggested structure might not be all that unstable. Note that 
there is a lot of supporting radiation pressure!}

\discuss{Kulkarni}{Angular momentum will only be efficiently removed from a mass-losing star 
if the core is coupled to the envelope. Could you comment on our current understanding of this coupling?}

\discuss{Vink}{There is a debate regarding the magnetic coupling of the core and the envelope. Some 
massive star evolution modellers include magnetic fields, which results in a strong core-envelope coupling
(e.g. Brott et al. 2011, A\&A 530, 115), as this seems 
to be favoured when regarding the spins of neutron stars 
(Langer/Bonn argument). The Geneva models do not include magnetic fields, leading to less coupling.}

\end{discussion}

\end{document}